\documentclass[pra,twocolumn,superscriptaddress,aps,showpacs]{revtex4-1}

\usepackage{physics}
\usepackage{amsmath}
\usepackage{amssymb}
\usepackage{graphicx}
\usepackage{float}
\graphicspath{{PNG/}}
\usepackage{xcolor}

\makeatletter
\newcommand*\bigcdot{\mathpalette\bigcdot@{.5}}
\newcommand*\bigcdot@[2]{\mathbin{\vcenter{\hbox{\scalebox{#2}{$\m@th#1\bullet$}}}}}
\makeatother

\begin{document}
\title{Two-body quench dynamics of harmonically trapped interacting particles}

\author{A. D. Kerin and A. M. Martin}
\affiliation{School of Physics, University of Melbourne, Parkville, VIC 3010, Australia}

\date{\today}


\begin{abstract}
We consider the quantum evolution of a pair of interacting atoms in a three dimensional isotropic trap where the interaction strength is quenched from one value to another. Using exact solutions of the static problem we are able to evaluate time-dependent observables such as the overlap between initial and final states and the expectation value of the separation between the two atoms. In the case where the interaction is quenched from the non-interacting regime to the strongly interacting regime, or vice versa, we are able to obtain analytic results. Examining the overlap between the initial and final states we show that when the interaction is quenched from the non-interacting to strongly interacting regimes the early time dependence dynamics are consistent with theoretical work in the single impurity many-body limit. When the system is quenched from the strongly to non-interacting regime we predict large oscillations in the separation between the two atoms, which arises from a logarithmic divergence due to the zero-range nature of the interaction potential.
\end{abstract}
\maketitle


\section{Introduction}
Understanding the evolution of non-equilibrium quantum states is pertinent to a wide range of systems, such as, superfluid turbulence, mesoscopic electrical circuits and even neutron stars. Such systems typically involve many interacting components and as such it is essential to make appropriate approximations when trying to determine their non-equilibrium properties. Alternatively it is possible to construct few-body systems experimentally and probe their non-equilibrium quantum phenomena. For example harmonically trapped quantum gases can be constructed atom by atom~\cite{serwane2011deterministic, murmann2015two, zurn2013pairing, zurn2012fermionization}. In such systems it is possible to precisely control the number of atoms and their interactions. In this context we focus on the quench dynamics of two atoms in a harmonic trap interacting via contact interactions.

The study of harmonically trapped few-body systems with contact interactions~\cite{busch1998two, fedorov2001regularization, PhysRevA.74.053604, werner2006unitary, PhysRevA.76.033611, PhysRevA.81.053615, PhysRevA.90.023626,daily2012occupation} has previously been used to gain insight into the thermodynamic properties of quantum gases~\cite{PhysRevLett.102.160401, liu2010three, PhysRevLett.96.030401, Cui2012, PhysRevA.85.033634, PhysRevLett.107.030601, PhysRevA.85.053636, PhysRevA.86.053631, Nature463_2010, Science335_2010,levinsen2017universality}, particularly in the strongly interacting regime, and have been experimentally studied in their own right~\cite{PhysRevLett.96.030401}. In this paper we focus using the solutions for two interacting atoms in a harmonic trap~\cite{busch1998two}, a regime which is experimentally achievable~\cite{murmann2015two, zurn2012fermionization}, to determine the quench dynamics of such a system. Specifically, we consider scenarios where the zero-range s-wave scattering length ($a_{\rm s}$) is quenched via Feshbach resonance~\cite{fano1935feshbackh, feshbach1958feshbackh, chin2010feshbach, tiesinga1993feshbackh} from one value to another. For such a system we determine the quantum dynamics of the state and evaluate the overlap between the initial and final states (Ramsey signal) and the expectation value of the separation between the atoms. There have been experimental measures of the Ramsey signal~\cite{cetina2016ultrafast} in the regime of impurity atoms residing in a Fermi sea of other atoms and there have been experimental measures of the separation between \textsuperscript{6}Li atoms in a quenched anisotropic trap~\cite{guan2019density}. This experimental work has  been complimented by theoretical work~\cite{PhysRevA.84.063632, PhysRevX.2.041020, parish2016quantum, Schmidt_2018, PhysRevLett.122.205301, PhysRevB.101.045134, bougas2020stationary, bougas2019analytical, budewig2019quench, kehrberger2018quantum, sykes2013quenching, corson2015bound, d2018efimov}. Our work is the trapped two-body limit of such systems. In this context we find that the evolution of the Ramsey signal aligns with equivalent many-body results in the short-time limit \cite{cetina2016ultrafast, parish2016quantum}, thus demonstrating that the early time many-body dynamics are governed by the two-body interactions as found in Refs. \cite{parish2016quantum,sykes2013quenching,corson2015bound} and are independent of trapping geometry. For the evaluation of the dynamics of the expectation value of the separation between the atoms, we find that when the initial state is strongly interacting and it is quenched to a non-interacting regime large oscillations are predicted. 

\section{Overview of the Two-Body Problem}
In this work we explore the quench dynamics of a two-body system residing in a harmonic potential. Specifically we consider two distinguishable particles with masses $m_{1}$ and $m_{2}$ at positions $\vec{r}_{1}$ and $\vec{r}_{2}$.
These particles interact via a contact interaction described by the Fermi pseudo-potential \cite{huang1957quantum}. The static properties of such a system can be described with the following two-body Hamiltonian
\begin{eqnarray}
\hat{H}_{\rm 2b}&=&-\frac{\hbar^2}{2M}\nabla^{2}_{R}-\frac{\hbar^2}{2\mu}\nabla^{2}_{r}
+\frac{M\omega^2R^2}{2}+\frac{\mu\omega^2r^2}{2}\nonumber\\
&+&\frac{2\pi\hbar^2 a_{s}}{\mu}\delta^3(r)\frac{\partial}{\partial r}(r \bigcdot),
\label{eq:2b}
\end{eqnarray}
where we have used the following co-ordinate transformations, 
\begin{eqnarray}
\vec{R}=\frac{m_{1}\vec{r}_{1}+m_{2}\vec{r}_{2}}{M}&, \quad \vec{r}=\vec{r}_{1}-\vec{r}_{2}. \nonumber
\end{eqnarray}
In the above $M=m_{1}+m_{2}$ and $\mu=m_{1}m_{2}/M$, $\omega$ is the harmonic trapping frequency and $a_{\rm s}$ is the s-wave scattering length which characterizes the strength of the contact interactions.
\\\\
From Eq.~(\ref{eq:2b}) it is clear the the two-body Hamiltonian can be split into two parts,
\begin{eqnarray}
\hat{H}_{{\rm CoM}}&=&-\frac{\hbar^2}{2M}\nabla^{2}_{R}+\frac{M\omega^2R^2}{2},\\
\hat{H}_{{\rm Rel}}&=&-\frac{\hbar^2}{2\mu}\nabla^{2}_{r}+\frac{\mu\omega^2r^2}{2}
+\frac{2\pi\hbar^2 a_{s}}{\mu}\delta^3(r)\frac{\partial}{\partial(r)}(r \bigcdot),
\end{eqnarray}
such that $\hat{H}_{\rm 2b}=\hat{H}_{{\rm CoM}}+\hat{H}_{{\rm Rel}}$.
\\\\
The solution to $\hat{H}_{{\rm CoM}}$ is a simple harmonic oscillator wavefunction $\phi(R)$. Additionally, the relative two-body wavefunction is well understood \cite{busch1998two}: 
\begin{eqnarray}
\psi(\nu,r)&=&N_{\nu}\Gamma(-\nu)e^{-\tilde{r}^2/2}U\left(-\nu,3/2,\tilde{r}^2\right)\\
&=& N_{\nu}e^{-\tilde{r}^2/2}\sum_{k=0}^{\infty}\frac{L_k^{\frac{1}{2}}(\tilde{r}^2)}{k-\nu},
\label{eq:psi1}
\end{eqnarray}
where $L_{k}^{\frac{1}{2}}({\tilde r}^2)$ are the associated Laguerre polynomials, $U(-\nu,3/2,\tilde{r}^2)$ is Kummer's function of the second kind, $\tilde{r}=r/a_{{\rm ho}}$ with $a_{{\rm ho}}=\sqrt{\hbar/\mu \omega }$, and $\nu$ is the energy pseudo-quantum number of the relative wavefunction such that, $E_{{\rm Rel}}=(2\nu+3/2)\hbar\omega$. The normalization, $N_{\nu}$, is given by $N_{\nu}=\left( 2\pi a_{ho}^3 Z(\nu) \right)^{-\frac{1}{2}}$ where
\begin{eqnarray}
Z(\nu)=\frac{\pi\Gamma(1-\nu)\left[\psi^{(0)}(-\nu-1/2)-\psi^{(0)}(-\nu)\right]}{\nu\Gamma(-\nu-1/2)},
\end{eqnarray}
with $\psi^{(0)}$ being the digamma function of degree $0$ \cite{abramowitz1964handbook}.

\begin{figure}
\includegraphics[height=5.5cm,width=8.5cm]{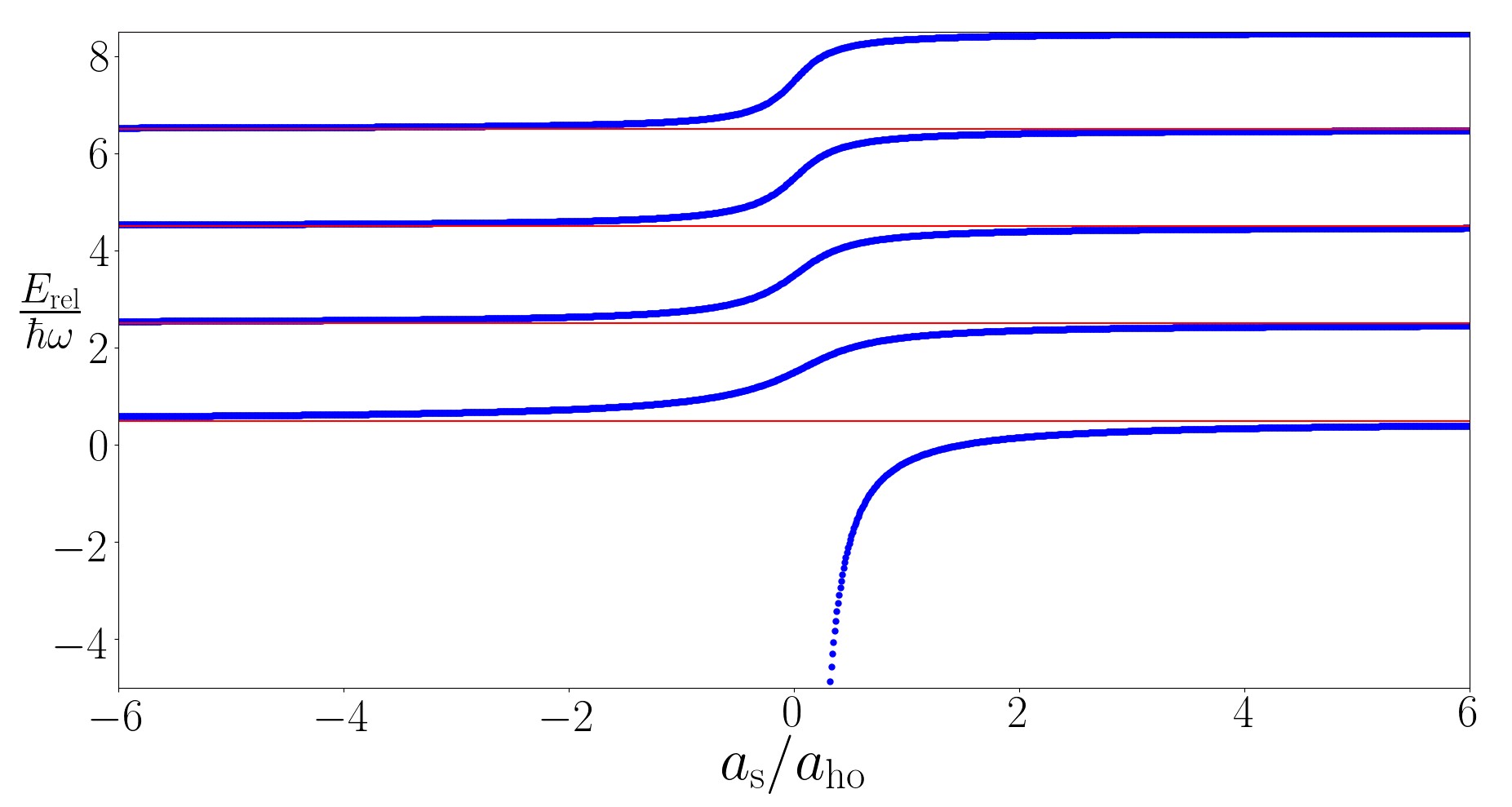}
\caption{Eigenenergies of $\hat{H}_{{\rm Rel}}$ as a function of the s-wave scattering length (blue dots). The red lines correspond to $E_{rel}=(2n+1/2)\hbar \omega$ with $n \in \mathbb{Z}_{\geq 0}$, the eigenenergy of the relative Hamiltonian in the unitary limit ($a_{\rm s}/a_{{\rm ho}}\rightarrow \pm \infty$).}
\label{fig:eigen}
\end{figure}

The values of the energy pseudo-quantum number, $\nu$, for a given interaction strength, $a_{\rm s}$, can be determined by the following transcendental equation \cite{busch1998two}
\begin{eqnarray}
\frac{a_{{\rm ho}}}{ a_{{\rm s}}}=\frac{2\Gamma(-\nu)}{\Gamma(-\nu-1/2)}. \label{eq: eigenstates}
\end{eqnarray}
The solutions to Eq.~(\ref{eq: eigenstates}) are plotted in Fig.~\ref{fig:eigen}. As one expects in the non-interacting limit, $a_{{\rm s}}/a_{{\rm ho}}\rightarrow0$, we recover the simple harmonic oscillator energy spectrum: $E_{{\rm Rel}}=(2n +3/2)\hbar \omega$ with $n \in \mathbb{Z}_{\geq0}$. In the strongly interacting (unitary) limit, $a_{\rm s}/a_{{\rm ho}}\rightarrow \pm \infty$, it is found that $E_{{\rm Rel}}=(2n +1/2)\hbar \omega$ with $n \in \mathbb{Z}_{\geq0}$ (solid horizontal red lines) i.e. $\nu \rightarrow n-1/2$.

\section{Quench Dynamics}
In general if a system is initially in a state $|\Phi \rangle$ and it is quenched then the state evolves as
\begin{eqnarray}
\ket{\Psi(t)}=e^{-i\hat{H^{\prime}}t/\hbar}\ket{\Phi}=\sum_{j=0}^{\infty}\bra{\Phi_{j}^{\prime}}\ket{\Phi}e^{-iE_{j}^{\prime}t/\hbar}\ket{\Phi_{j}^{\prime}},
\end{eqnarray}
where $\Phi_{j}^{\prime}$ are the eigenstates of the post-quench Hamiltonian $\hat{H}^{\prime}$, with corresponding eigenenergies $E_{j}^{\prime}$ and the initial pre-quench eigenstate, associated with the Hamiltonian $\hat{H}$, is $\Phi$. 

In the context of the two-body system we are considering this implies
\begin{eqnarray}
\ket{\Psi(t)}&=&\sum_{j,k=0}^{\infty}\bra{\phi_{k}^{\prime}(R)\psi_j^{\prime}(r)}\ket{\phi(R)\psi(r)} \nonumber \\
&\times&  e^{-i\left(E_{j}^{\prime}+E_{k}^{\prime}\right)t/\hbar}\ket{\phi_{k}^{\prime}(R)\psi_j^{\prime}(r)}. \label{eq:psi_t_}
\end{eqnarray}
Since the quenches considered in this work are in the s-wave scattering length the center of mass eigenstates remain unaffected and Eq.~(\ref{eq:psi_t_}) reduces to 
\begin{eqnarray} 
\ket{\Psi(t)}&=&\sum_{j=0}^{\infty}\bra{\psi_j^{\prime}(r)}\ket{\psi(r)} \nonumber \\
&\times& e^{-i\left(E^{\prime}_{j}+E_{\rm CoM}\right)t/\hbar}\ket{\phi(R)\psi_j^{\prime}(r)}. \label{eq:psi_t_2}
\end{eqnarray}

Following this formalism we will calculate two observables. In each case we start from an initial state $\ket{\Phi}=\ket{\phi(R)\psi(r)}$ characterized by the s-wave scattering length, $a_{\rm s}$. The system is then quenched to a new s-wave scattering length $a_{\rm s}^{\prime}$. The two observables we consider are the Ramsey signal, 
\begin{eqnarray}
S(t) =\bra{\Phi}\ket{\Psi(t)},
\end{eqnarray}
the magnitude of which is the contrast, and the separation between the particles, 
\begin{eqnarray}
\langle r(t) \rangle=\bra{\Psi(t)}r\ket{\Psi(t)}.
\end{eqnarray}

\subsection{Contrast}
After a quench the dynamics of the system can be probed through a Ramsey interferometric measurement \cite{cetina2016ultrafast, PhysRevA.84.063632, PhysRevX.2.041020, parish2016quantum}, which allows the contrast,
to be measured. Previous work regarding the contrast (and other quantities) in one- and two-dimensional systems of two bosons has been undertaken \cite{bougas2020stationary,bougas2019analytical,budewig2019quench, kehrberger2018quantum}.

For our three-dimension system, utilising Eq.~(\ref{eq:psi_t_2}), the Ramsey signal reduces to 
\begin{eqnarray}
S(t)=\sum_{j=0}^{\infty}\left| \bra{\psi(r)}\ket{\psi_{j}^{\prime}(r)}\right|^2e^{-i\left(E^{\prime}_{j}-E_{{\rm Rel}}\right)t/\hbar}, \label{eq:S}
\end{eqnarray}
where $\ket{\psi(r)}$ is the initial eigenstate, with eigenenergy  $E_{{\rm Rel}}$, of $\hat{H}_{{\rm Rel}}$, with an s-wave scattering length $a_{\rm s}$ and the states $\ket{\psi_{j}^{\prime}(r)}$ are the eigenstates of $\hat{H}^{\prime}_{{\rm Rel}}$, with an s-wave-scattering length $a_{\rm s}^{\prime}$, with corresponding eigenenergies $E_{j}^{\prime}$. The general properties of Eq.~(\ref{eq:S}) can be understood via the realization that $\ket{\psi_{j}^{\prime}(r)}$ rotates at an angular frequency  $E_{j}^{\prime}$ and $\left| \bra{\psi(r)}\ket{\psi_{j}^{\prime}(r)}\right|^2$ determines the magnitude of the squared overlap of the pre-quench state and the $j^{th}$ post-quench eigenstate.

In general the coefficients, $\left| \bra{\psi(r)}\ket{\psi_{j}^{\prime}(r)}\right|^2$, of the sum in Eq.~(\ref{eq:S})  can be evaluated. Specifically, defining $\nu$ to be the pseudo-quantum number for the initial state, $\ket{\psi(r)}$, and $\nu_j^{\prime}$ to be the energy pseudo-quantum numbers for the final states, $\ket{\psi_{j}^{\prime}(r)}$, then
\begin{eqnarray}
\bra{\psi(r)}\ket{\psi_{j}^{\prime}(r)}&=&2\pi N_{\nu}N_{\nu_{j}^{\prime}}a_{{\rm ho}}^3\sum_{k=0}^{\infty} \frac{\Gamma(k+3/2)}{(k-\nu)(k-\nu_{j}^{\prime})\Gamma(k+1)}\nonumber\\
&=&\frac{\sqrt{\pi}}{2 \nu \nu_{j}^{\prime}}\frac{{}_{3}F_{2}\left(\frac{3}{2},-\nu_{j}^{\prime},-\nu;1-\nu_{j}^{\prime},1-\nu;1\right)}{\sqrt{Z(\nu)Z(\nu_{j}^{\prime})}} \nonumber \\
\label{eq:coefs}
\end{eqnarray}
where ${}_{3}F_{2}(a,b,c;d,e;f)$ is a generalized hypergeometric function. From Eqs.~(\ref{eq: eigenstates},\ref{eq:S},\ref{eq:coefs}) it is possible to numerically evaluate $S(t)$ for any quench from $a_{\rm s}$ to $ a_{\rm s}^{\prime}$.

Below we consider two scenarios where the s-wave scattering length is quenched from $a_{\rm s}/a_{{\rm ho}}=0$ to $a_{\rm s}^{\prime}/a_{{\rm ho}}=\pm \infty$ and from $a_{\rm s}/a_{{\rm ho}} = \pm \infty$ to $a_{\rm s}^{\prime}/a_{{\rm ho}}=0$. In these cases exact closed form expressions for $S(t)$ can be found. From Eq.~(\ref{eq: eigenstates}), and as can be seen in Fig.~\ref{fig:eigen}, in the unitary limit $\nu_{j}\rightarrow j-1/2$. In the non-interacting to unitary case the initial state, $\ket{\psi(r)}$, is the simple harmonic wavefunction with $l=m=0$:
\begin{eqnarray}
\psi_{n00}(r)&=&N_{n0}e^{-\tilde{r}^2/2}L_{n}^{\frac{1}{2}}(\tilde{r}^2)Y_{00}(\theta,\phi),\label{eq:SHO} \\
N_{n0}&=&\sqrt{\sqrt{\frac{1}{4\pi a_{\rm ho}^6}}\frac{2^{n+3}n!}{(2n+1)!!}}, \nonumber
\end{eqnarray}
where $Y_{00}(\theta,\phi)$ is the lowest order spherical harmonic.

\begin{figure}
\includegraphics[height=5.5cm,width=8.5cm]{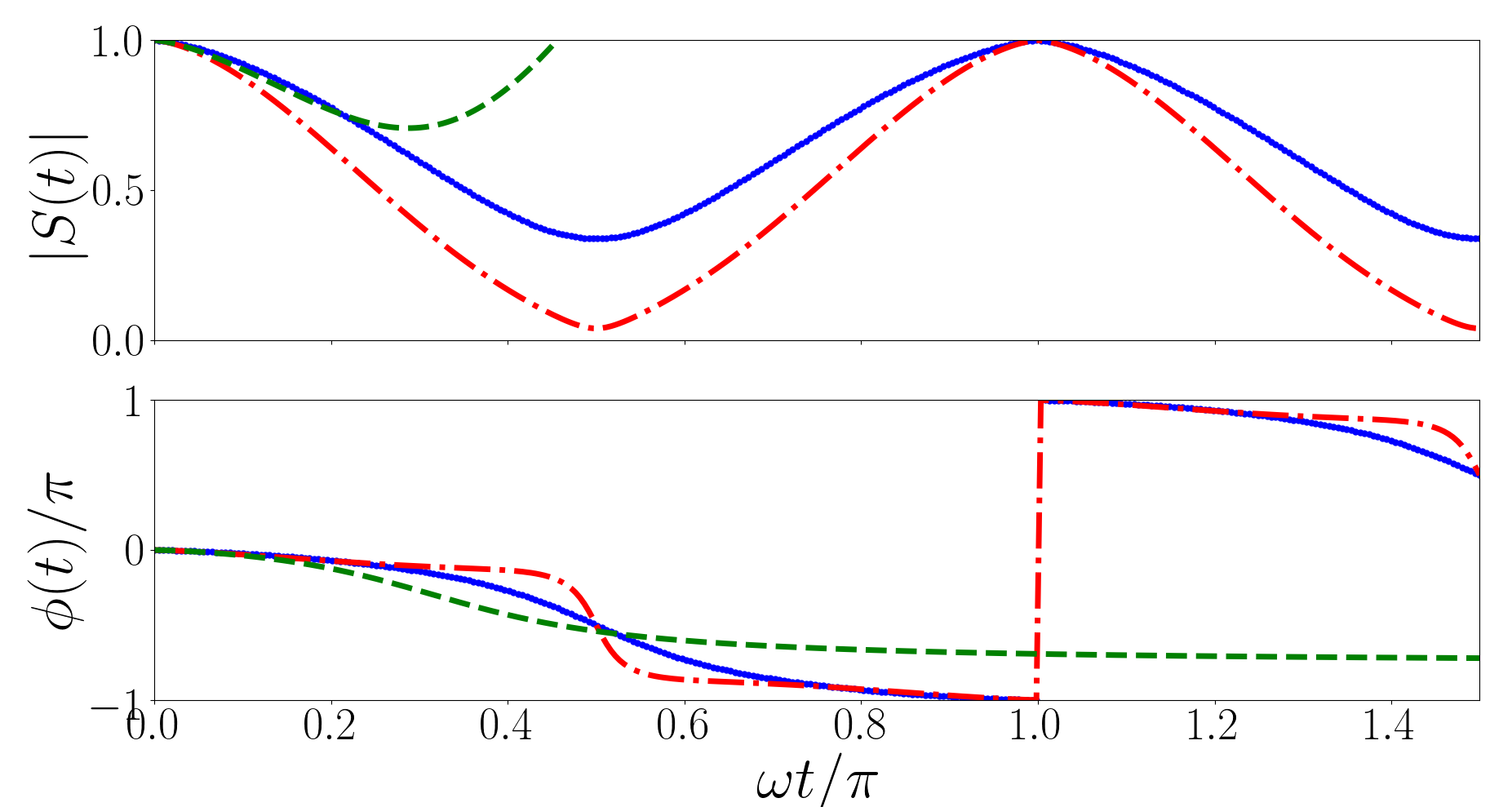}
\caption{Ramsey signal for $a_{\rm s}/a_{{\rm ho}}=0$ to $a_{\rm s}^{\prime}/a_{{\rm ho}}=\infty$ case as a function of $t$. The upper (lower) panel plots $|S(t)|$ ($\phi(t)$), where the Ramsey signal has been parameterized as $S(t)=|S(t)|e^{-i\phi(t)}$. The blue solid curve corresponds to the exact ground state result, Eq.~(\ref{eq:S_n0}), the dashed-dotted red curve corresponds to the exact $n=1$ result of Eq.~(\ref{eq:S_full}), and the dashed green curve corresponds to the early time expansion for the ground state result, Eq.~(\ref{eq:S_early}).}
\label{fig:S1}
\end{figure}

For the case where $a_{\rm s}/a_{{\rm ho}}=0$ and $a_{\rm s}^{\prime}/a_{{\rm ho}}=\pm \infty$ and the initial state has quantum number $n$,
\begin{eqnarray}
S(t)&=&\frac{4e^{i(2n+1)\omega t}\Gamma(n+3/2)}{(1+2n)^2 \pi^\frac{3}{2}\Gamma(1+n)} \nonumber \\
&\times&{}_{3}F_{2}\left(\frac{1}{2},-\frac{1}{2}-n,-\frac{1}{2}-n;\frac{1}{2}-n,\frac{1}{2}-n;e^{-2i\omega t}\right). \nonumber \\
\label{eq:S_full}
\end{eqnarray}
In the case where the initial state is the ground state ($n=0$) this reduces to
\begin{eqnarray}
S(t)=\frac{2}{\pi} \left[e^{i\omega t}\sqrt{(1-e^{-i2\omega t})}+\arcsin(e^{-i\omega t})\right].
\label{eq:S_n0}
\end{eqnarray} 
The contrast is plotted in Fig.~\ref{fig:S1} (solid blue curve for Eq.~(\ref{eq:S_n0}) and dashed-dotted red curve for Eq.~(\ref{eq:S_full}) with $n=1$), with the upper figure showing the magnitude, $|S(t)|$, and lower figure showing the phase, $\phi(t)$, where we have parameterized the Ramsey signal as $S(t)=|S(t)|e^{-i\phi(t)}$. In general the contrast oscillates, with period $\pi/\omega$, whilst the phase exhibits a period of $2\pi/\omega$ with a {\it stepping} behaviour ($\omega t\approx \pi/2$) that is particularly pronounced in the $n\geq 1$ case. A similar phase feature has been experimentally and theoretically observed for the case of a single impurity in a uniform Fermi sea \cite{cetina2016ultrafast}.

It is also informative to examine the short time dynamics of the contrast in the limit $\omega t \ll 1$. Expanding Eq.~(\ref{eq:S_n0}) we find
\begin{eqnarray}
S(t)\approx 1 - \frac{11}{3\sqrt{2}\pi}e^{-i \pi/4}(\omega t)^{3/2},
\label{eq:S_early}
\end{eqnarray}
which is also plotted in Fig.~\ref{fig:S1} (dashed green curve).
\\\\
Previous calculations which perturbatively evaluate the Ramsey signal for a single Fermi impurity in a uniform sea of fermions provide a useful comparison for the exact two-body results presented above. For such a case, where the interaction is quenched from $a_{\rm s}/a_{{\rm ho}}=0$ to $a_{\rm s}^{\prime}/a_{{\rm ho}}=\pm \infty$ it has been shown that the short time dynamics follow \cite{parish2016quantum}
\begin{eqnarray}
S(t)\approx 1 - \frac{8}{9}\left(\frac{3m_1}{2\pi\mu}\right)^{3/2}e^{-i \pi/4}(\omega t)^{3/2},
\label{eq:S_early_many}
\end{eqnarray}
where we have assumed that the Fermi energy is $3 \hbar \omega/2$. 

Equations (\ref{eq:S_early}) and (\ref{eq:S_early_many}) show that the short time dynamics are dominated by $1-\alpha t^{3/2}$ in each case. Perhaps what is more remarkable is that for $m_1=m_2$ we find that for perturbative calculation, Eq.~(\ref{eq:S_early_many}), $\alpha=0.829\dots e^{-i \pi/4}\omega^{3/2}$ as compared to $\alpha=0.825\dots e^{-i \pi/4}\omega^{3/2}$ for the two-body calculation, Eq.~(\ref{eq:S_early}). This supports the idea that the early time dynamics are dominated by two-body effects \cite{sykes2013quenching} and are independent of the trapping potential \cite{corson2015bound}.

\begin{figure}
\includegraphics[height=5.5cm,width=8.5cm]{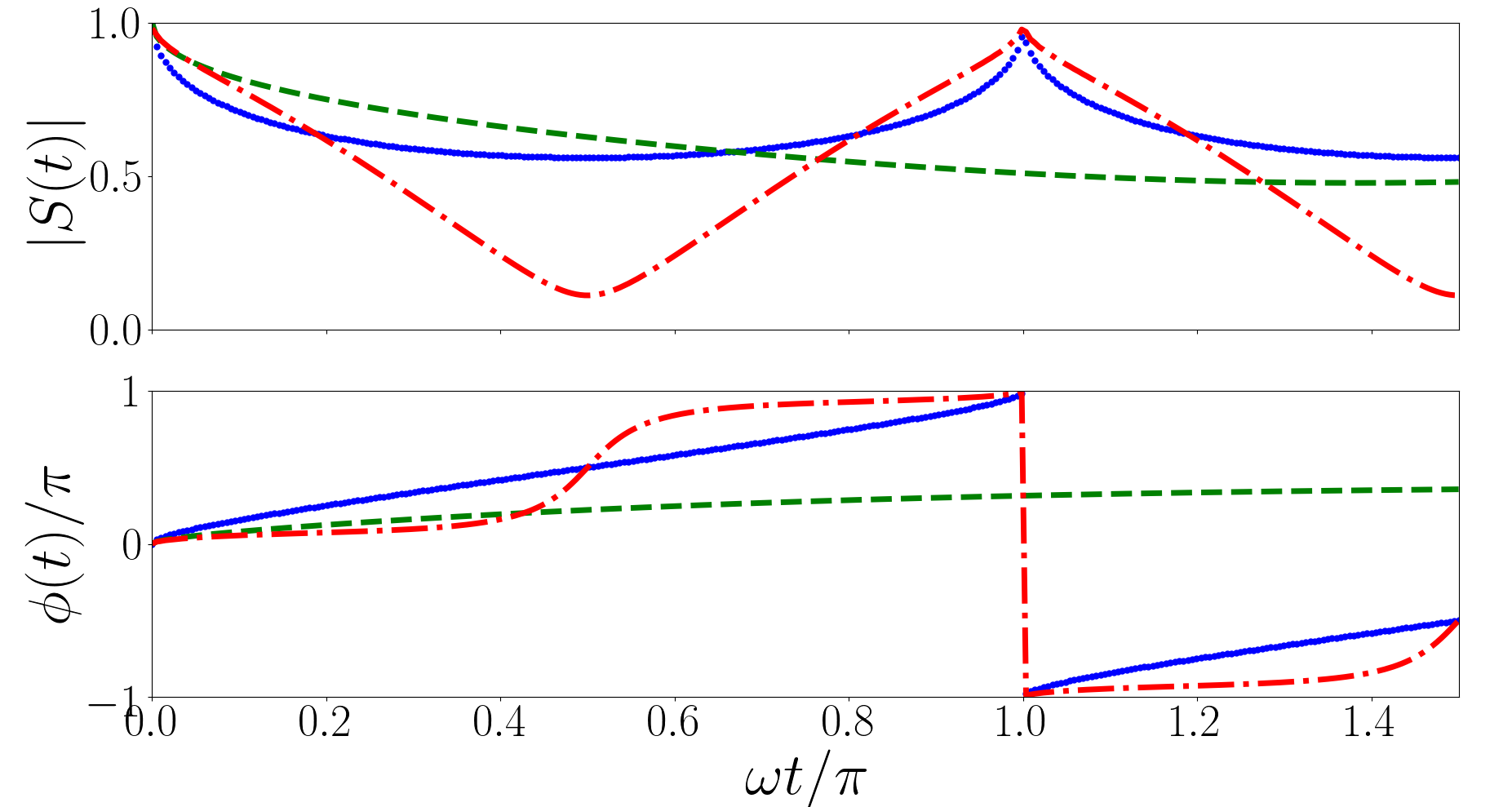}
\caption{Ramsey signal for $a_{\rm s}/a_{{\rm ho}}=\infty$ and $a_{\rm s}^{\prime}/a_{{\rm ho}}=0$  as a function of $t$. The upper (lower) panel plots $|S(t)|$ ($\phi(t)$), where the Ramsey signal has been parameterized as $S(t)=|S(t)|e^{-i\phi(t)}$. The blue solid curve corresponds to the exact result, Eq.~(\ref{eq:S2_n0}), the dashed-dotted red curve corresponds to the exact $j=1$ result of Eq.~(\ref{eq:S2_full}), and the dashed green curve corresponds to the early time expansion of the ground state result Eq.~(\ref{eq:S2_early}).}
\label{fig:S2}
\end{figure}

We also consider the reverse case where the system is initially in the unitary limit and quenched to the non-interacting limit. In the unitary case ($a_{\rm s}/a_{{\rm ho}} = \pm \infty$) with pseudo-quantum number $\nu_{j}=j-1/2$:
\begin{eqnarray}
\psi(\nu_{j},r)=N_{j-1/2}e^{-\tilde{r}^2/2}\sum_{n=0}^{\infty}\frac{L_{n}^{\frac{1}{2}}(\tilde{r}^2)}{n-j+1/2},
\end{eqnarray}
where
\begin{eqnarray}
N_{j-1/2}=\sqrt{\frac{\Gamma(1/2+j)}{2\pi^3 a_{\rm ho}^3 \Gamma(1+j)}}. \nonumber
\end{eqnarray}
The final states are in the non-interacting limit ($a_{\rm s}^{\prime}/a_{{\rm ho}}=0$), and so are simple harmonic wavefunctions as defined in Eq.~(\ref{eq:SHO}). As a result the Ramsey signal is evaluated as
\begin{eqnarray}
S(t)&=&\frac{2e^{-i(1-2j)}\Gamma(j+1/2)}{(2j-1)^2 \pi^\frac{3}{2}\Gamma(j+1)}\nonumber\\
&\times& {}_{3}F_{2}\left(\frac{3}{2},\frac{1}{2}-j.\frac{1}{2}-j;\frac{3}{2}-j,\frac{3}{2}-j;e^{-2i\omega t}\right), \nonumber \\
\label{eq:S2_full}
\end{eqnarray}
where $j$ denotes the excitation of the initial state.
In the case where the initial state is the ground state ($j=0$) the Ramsey signal reduces to
\begin{eqnarray}
S(t)=\frac{2}{\pi}\arcsin(e^{-i\omega t}).
\label{eq:S2_n0}
\end{eqnarray}
For early times ($\omega t \ll 1$) this becomes 
\begin{eqnarray}
S(t)\approx 1 - \frac{2\sqrt{2}}{\pi}e^{-i \pi/4}\left(1+\frac{e^{i \pi/2}}{12}\omega t\right)(\omega t)^{1/2}.
\label{eq:S2_early}
\end{eqnarray}
The exact ground state result, Eq.~(\ref{eq:S2_n0}) (solid blue curve), the early time dynamics, Eq.~(\ref{eq:S2_early}) (dashed green curve), and the exact first excited state result, Eq.~(\ref{eq:S2_full}) with $j=1$ (dashed-dotted red curve), are plotted in Fig.~\ref{fig:S2}. In general the contrast oscillates, with period $\pi/\omega$, whilst the phase exhibits a period of $2\pi/\omega$ with a {\it stepping} behaviour ($\omega t\approx \pi/2$) that is again particularly pronounced in the $n\geq 1$ case. In contrast to the previous case, shown in Fig.~\ref{fig:S1}, we find that when the system is quenched from unitarity to non-interacting $S(t)$ exhibits  non-analytic cusps at $\omega t= 0,\pi,2\pi\dots$. 

The Ramsey signal has been experimentally measured using Ramsey interferometry techniques \cite{cetina2016ultrafast}. As outlined by \cite{PhysRevX.2.041020, parish2016quantum} the Ramsey signal can be related to the difference in probabilities of different states.

Consider the two particles $1$ and $2$. Suppose that particle $1$ has only one allowable state, $\ket{1}$, and particle $2$ has two: $\ket{2\uparrow}$ and $\ket{2\downarrow}$. We assume that $\ket{2\downarrow}$ does not interact with $\ket{1}$ but $\ket{2\uparrow}$ can, the interactions are parameterized by $a_{\rm s}$. 
\\
For a general initial state of the system
\begin{eqnarray}
\ket{\Phi}=(a\ket{2\uparrow}+b\ket{2\downarrow}) \otimes \ket{1}.
\end{eqnarray}
For such a state a Ramsey pulse sequence has the following form:
\begin{eqnarray}
\ket{\Psi(t)}=U_{\pi/2}^{\phi}T(t)U_{\pi/2}^{\phi=0}
\ket{\Phi},
\end{eqnarray}
where
\[
U_{\pi/2}^{\varphi}=\frac{1}{\sqrt{2}}
\begin{bmatrix}
1 & -ie^{i\varphi}\\
-ie^{-i\varphi} & 1\\
\end{bmatrix},
\quad
T(t)=
\begin{bmatrix}
e^{-i\hat{H}^{\prime}t/\hbar} & 0\\
0 & e^{-i\hat{H}t/\hbar}\\
\end{bmatrix},
\]
and $\phi$ is the phase of the second $\pi/2$ pulse relative to first. Evaluating the probability of particle $2$ being in the $\ket{2\uparrow}$ ($P_{\uparrow}$) and $\ket{2\downarrow}$ ($P_{\downarrow}$) states given that the quench is applied after the first $\pi/2$ pulse results in
\begin{eqnarray}
{\rm P}_{\uparrow}&=&\frac{1}{2}\bra{\Psi(0)}( (a^2+b^2)+(b^2-a^2)\Re[e^{-i\phi}e^{-i(\hat{H}^{\prime}-\hat{H})t/\hbar}]\nonumber\\
&-&2ab\Im[e^{-i\phi}e^{-i(\hat{H}^{\prime}-\hat{H})t/\hbar}] )\ket{\Psi(0)}\\
{\rm P}_{\downarrow}&=&\frac{1}{2}\bra{\Psi(0)}( (a^2+b^2)+(a^2-b^2)\Re[e^{-i\phi}e^{-i(\hat{H}^{\prime}-\hat{H})t/\hbar}]\nonumber\\
&+&2ab\Im[e^{-i\phi}e^{-i(\hat{H}^{\prime}-\hat{H})t/\hbar}] )\ket{\Psi(0)}
\end{eqnarray}
and
\begin{eqnarray}
{\rm P}_{\uparrow}-{\rm P}_{\downarrow}&=&(b^2-a^2)\Re[e^{-i\phi}S(t)]-2ab\Im[e^{-i\phi}S(t)]. \nonumber \\
\end{eqnarray}
Hence we can connect an experimentally measurable quantity, ${\rm P}_{\uparrow}-{\rm P}_{\downarrow}$, to the Ramsey signal.
In all the observables calculated in this paper we have assumed $a=1$ and $b=0$.

\subsection{Particle Separation Expectation Value}
Experiment has shown that it is possible to measure the separation of two trapped \textsuperscript{6}Li atoms following a quench in trap geometry \cite{guan2019density} we believe it is possible to extend this to a quench in interaction strength.

Following the same methodology as presented in the previous section it is possible to evaluate the expectation value of the inter-particle separation via,
\begin{eqnarray}
\langle r (t) \rangle&=&\bra{\Psi(t)}r\ket{\Psi(t)}\nonumber\\
&=&\sum_{j,k=0}^{\infty}\bra{\psi_{n}(r)}\ket{\psi_j^{\prime}(r)}  \bra{\psi_k^{\prime}(r)}\ket{\psi_{n}(r)}\nonumber\\
&\times& \bra{\psi_j^{\prime}(r)}r\ket{\psi_k^{\prime}(r)} e^{-i(E_{k}^{\prime}-E_{j}^{\prime})t/\hbar}. \label{eq:ExpectR}
\end{eqnarray}  
We first consider the non-interacting $(a_{\rm s}/a_{\rm ho}=0)$ to interacting $(a^{\prime}_{\rm s}/a_{\rm ho}\neq0)$ case, where $\psi_{n}(r)$ is the non-interacting simple harmonic oscillator wavefunction, Eq.~(\ref{eq:SHO}), and $\psi^{\prime}_{j}(r)$ is the interacting wavefunction, Eq.~(\ref{eq:psi1}), with pseudo-quantum number $\nu_{j}$. In this case
\begin{eqnarray}
\bra{\psi_{n}(r)}\ket{\psi_j^{\prime}(r)}&=&\sqrt{\frac{n+1/2}{(n-\nu_{j})^2 Z(\nu_{j})}}\sqrt{\frac{\Gamma(n+1/2)}{\Gamma(n+1)}}, \nonumber \\
\end{eqnarray}
and
\begin{eqnarray}
\bra{\psi^{\prime}_{k}(r)}r\ket{\psi^{\prime}_{j}(r)}&=&N_{\nu_{j}}N_{\nu_{k}}\int re^{-\tilde{r}^2}\Gamma(-\nu_{j}) U(-\nu_{j},3/2,\tilde{r}^2)  \nonumber\\
&\times& \Gamma(-\nu_{k}) U(-\nu_{k},3/2,\tilde{r}^2) d^3 r\nonumber\\
&=&\frac{a}{\pi^2}\sqrt{\frac{\Gamma(j+\frac{1}{2})\Gamma(k+\frac{1}{2})}{\Gamma(j+1)\Gamma(k+1)}}\sum_{m=0}^{\infty}\sum_{n=0}^{\infty} \bigg[ \nonumber\\
& & \frac{(-1)^{m+n}}{(m-\nu_{j})(n-\nu_{k})} \binom{m+1/2}{n}\binom{n+1/2}{m} \bigg]\nonumber\\\label{eq:roverlap}
\end{eqnarray}
\begin{figure}
\includegraphics[height=5.5cm,width=8.5cm]{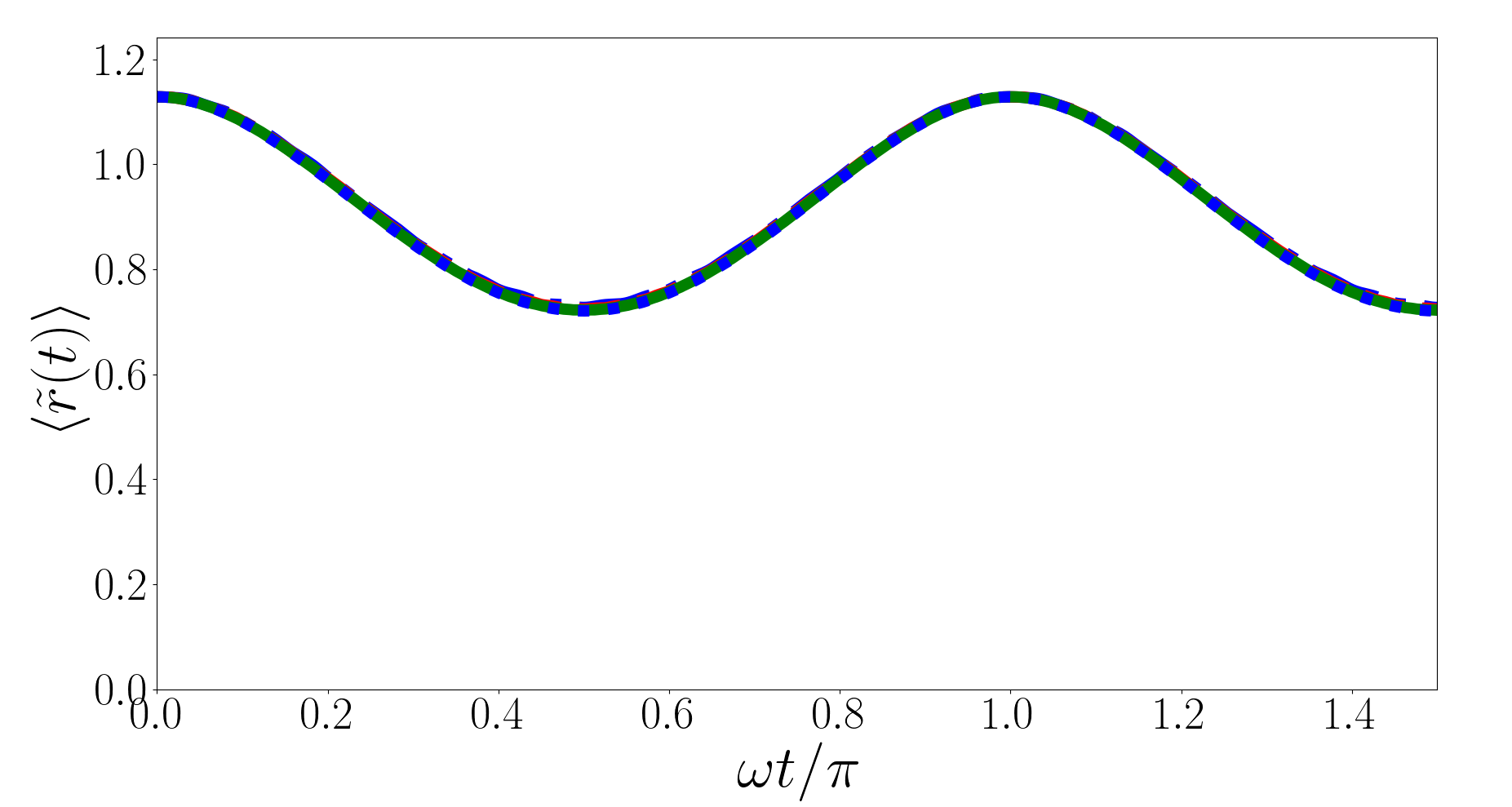}
\caption{The expectation of the particle separation as a function of time, where $a_{\rm s}/a_{\rm ho}=0$ and $a^{\prime}_{\rm s}/a_{\rm ho}=\infty$, for the $n=0$ case, Eq.~(\ref{eq:rt1}). Each curve corresponds to a different cut-off ($N_{\rm max}$) in the sum over $m$ and $l$ in Eq.~(\ref{eq:rt1}). Specifically, the  blue dotted curve corresponds to $N_{\rm max}=160$, the green solid curve corresponds to $N_{\rm max}=80$, the red dashed curve corresponds to  $N_{\rm max}=40$ and the dot-dashed blue curve corresponds to $N_{\rm max}=20$. Because this sum is convergent all curves are overlaid on top of one another.}
\label{fig:rt1}
\end{figure}
Note that by combining Eqs. (\ref{eq:coefs},~\ref{eq:ExpectR},~\ref{eq:roverlap}) one can calculate the expectation value of the particle separation for the general quench case.
In the case where the initial state is the ground state $(n=0)$ and the system is quenched to unitarity ($a^{\prime}_{\rm s}/a_{\rm ho}\rightarrow \infty$ and $\nu_{j}\rightarrow j-1/2$) we find
\begin{eqnarray}
\langle \tilde{r}(t) \rangle &=& -\frac{4}{\pi^{\frac{5}{2}}}\sum_{m,l=0}^{\infty} \frac{(-1)^{m+l}}{1+2m}e^{-i(1+2l)\omega t} \nonumber \\
&\times& \binom{\frac{1}{2}+m}{l}\binom{\frac{1}{2}+l}{m} \beta\left(e^{-2i \omega t},-\frac{1}{2}-l,\frac{3}{2}\right) \nonumber \\
&\times& {}_{2}F_{1}\left(-\frac{1}{2},-\frac{1}{2}-m,\frac{1}{2}-m,e^{2i\omega t}\right),
\label{eq:rt1}
\end{eqnarray}
where $\beta(z,a,b)$ is the incomplete beta function:
\begin{eqnarray}
\beta(z,a,b)=\int^{z}_{0} t^{a-1}(1-t)^{b-1}dt. \nonumber \\
\end{eqnarray}
Evaluating $\langle \tilde{r}(t) \rangle$ in this case results in an oscillating form, with period $\pi/\omega$, as shown in Fig.~\ref{fig:rt1}.

\begin{figure}
\includegraphics[height=5.5cm,width=8.5cm]{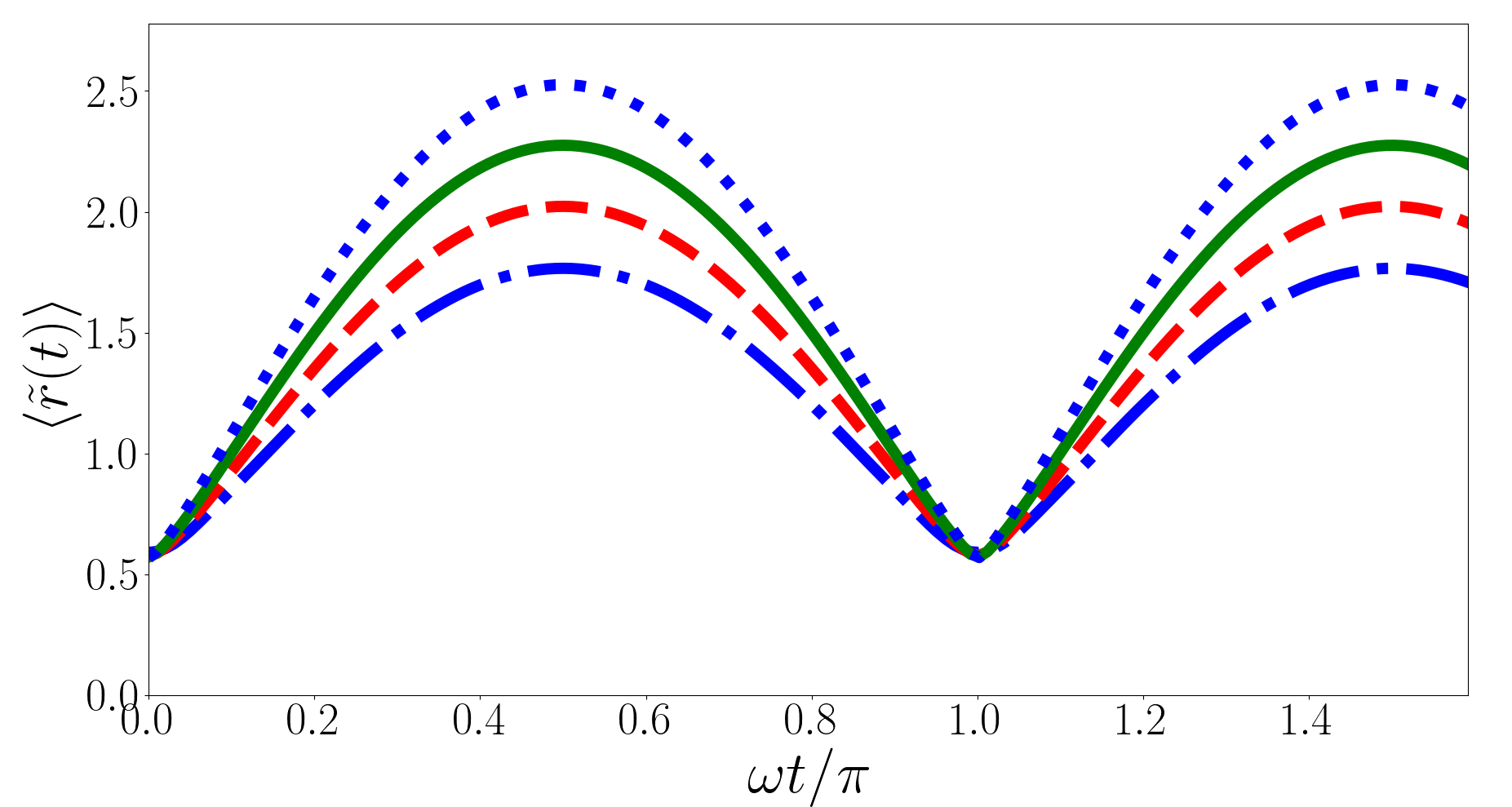}
\caption{The expectation of the particle separation as a function of time, where $a_{\rm s}/a_{\rm ho}=\infty$ and $a^{\prime}_{\rm s}/a_{\rm ho}=0$, for the $j=0$ case, Eq.~(\ref{eq:divergent}). Each curve corresponds to a different cut-off ($N_{\rm max}$) in the sum over $m$ and $l$ in Eq.~(\ref{eq:divergent}). Specifically, the  dotted blue curve corresponds to $N_{\rm max}=80$, the solid green curve corresponds to $N_{\rm max}=40$, the dashed red curve corresponds to  $N_{\rm max}=20$ and the blue dot-dashed curve corresponds to $N_{\rm max}=10$.}
\label{fig:r2}
\end{figure}

\begin{figure}
\includegraphics[height=5.5cm,width=8.5cm]{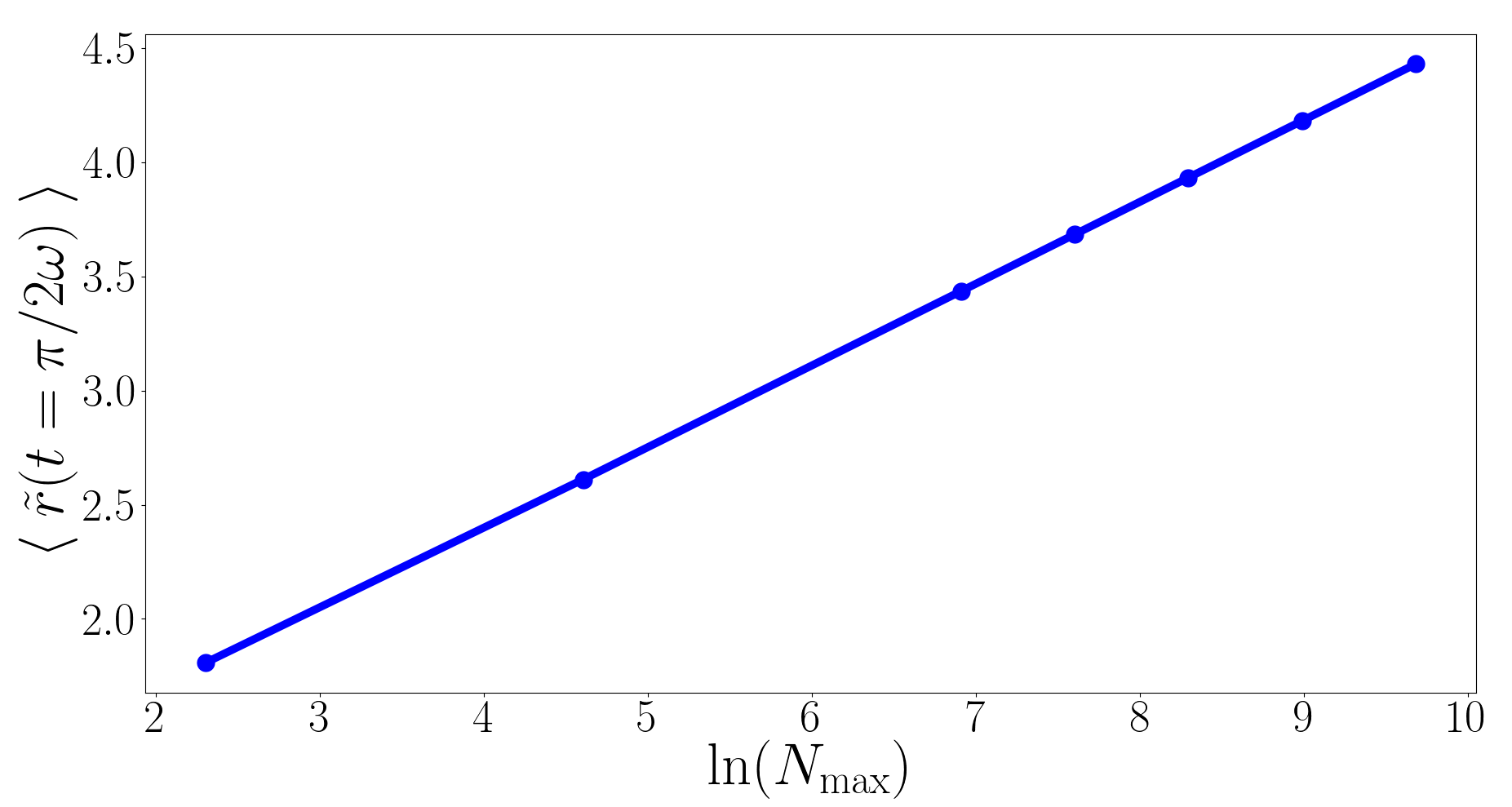}
\caption{$\langle \tilde{r}(t=\pi/2\omega) \rangle$ for the $j=0$ case as a function of the cut-off ($N_{\rm max}$) in the sum over $m$ and $l$ in Eq.~(\ref{eq:divergent}).}
\label{fig:r2_log}
\end{figure}

We now turn our attention to the case where the initial state is in the unitary limit $(a_{\rm s}/a_{\rm ho}=\infty)$ and the quench is to the non-interacting limit $(a^{\prime}_{\rm s}/a_{\rm ho}=0)$ where we find
\begin{eqnarray}
\langle \tilde{r}(t) \rangle &=& -\sum_{m,l=0}^{\infty} \frac{16 e^{-2i(m-l)\omega t}}{\Gamma(1+j)\Gamma(1+l)\Gamma(1+m)} \nonumber \\
&\times& \frac{\Gamma(1/2+j)\Gamma(3/2+m)\Gamma(3/2+l)}{(-1+4(m-l)^2)(1-2j+2m)(1-2j+2l)},\nonumber \\
\label{eq:divergent}
\end{eqnarray}
where $j$ denotes the initial state in the unitary limit. Eq.~(\ref{eq:divergent}) is logarithmically divergent in the summation. This divergence is most clearly demonstrated in Figs.~\ref{fig:r2} and \ref{fig:r2_log}. Fig.~\ref{fig:r2} shows, for $j=0$ that although $\langle \tilde{r}(t) \rangle$ is periodic, with period $\pi/\omega$ and the sum converges for $t\omega = k\pi$, where $k$ is an integer, it appears to diverge away from these points. Examining $\langle \tilde{r}(t=\pi/2\omega) \rangle$ elucidates this more clearly as demonstrated in Fig.~\ref{fig:r2_log}, which plots  $\langle \tilde{r}(t=\pi/2\omega) \rangle$  as a function of the maximum number of terms ($N_{\rm max}$) in the sum in Eq.~(\ref{eq:divergent}). As can be seen $\langle \tilde{r}(t=\pi/2\omega) \rangle$ diverges logarithmically.

The origins of this divergence can be understood by examining 
\begin{eqnarray}
I_{n}(r,t,N_{\rm max})&=&\sum_{l,m=0}^{N_{\rm max}}\bra{\psi_{n}(r)}\ket{\psi_l^{\prime}(r)}  \bra{\psi_m^{\prime}(r)}\ket{\psi_{n}(r)}\nonumber\\
&\times& \psi_l^{\prime}(r)r^{3}\psi_m^{\prime}(r)e^{-2i(m-l)\omega t},
\end{eqnarray} 
where for given initial state ($n$) 
\begin{eqnarray}
\langle \tilde{r}(t) \rangle =  4\pi \int_{0}^{\infty}   \lim_{N_{\rm max} \rightarrow \infty} I_{n}(r,t,N_{\rm max}) dr.
\end{eqnarray}
In the ground state ($n=0$) this reduces to,
\begin{eqnarray}
I_{0}({\tilde r},t,N_{\rm max})&=& \frac{8\tilde{r}^3 e^{-\tilde{r}^2}}{\pi^{\frac{5}{2}}}\sum_{l,m=0}^{N_{\rm max}}\frac{e^{-2i(m-l)\omega t}}{\Gamma(1+l)\Gamma(1+m)} \nonumber \\
&\times& \Gamma(1/2+l) {}_{1}F_{1}(-l,3/2,\tilde{r}^2) \nonumber \\
&\times&\Gamma(1/2+m){}_{1}F_{1}(-m,3/2,\tilde{r}^2).
\end{eqnarray}

\begin{figure}
\includegraphics[height=5.5cm,width=8.5cm]{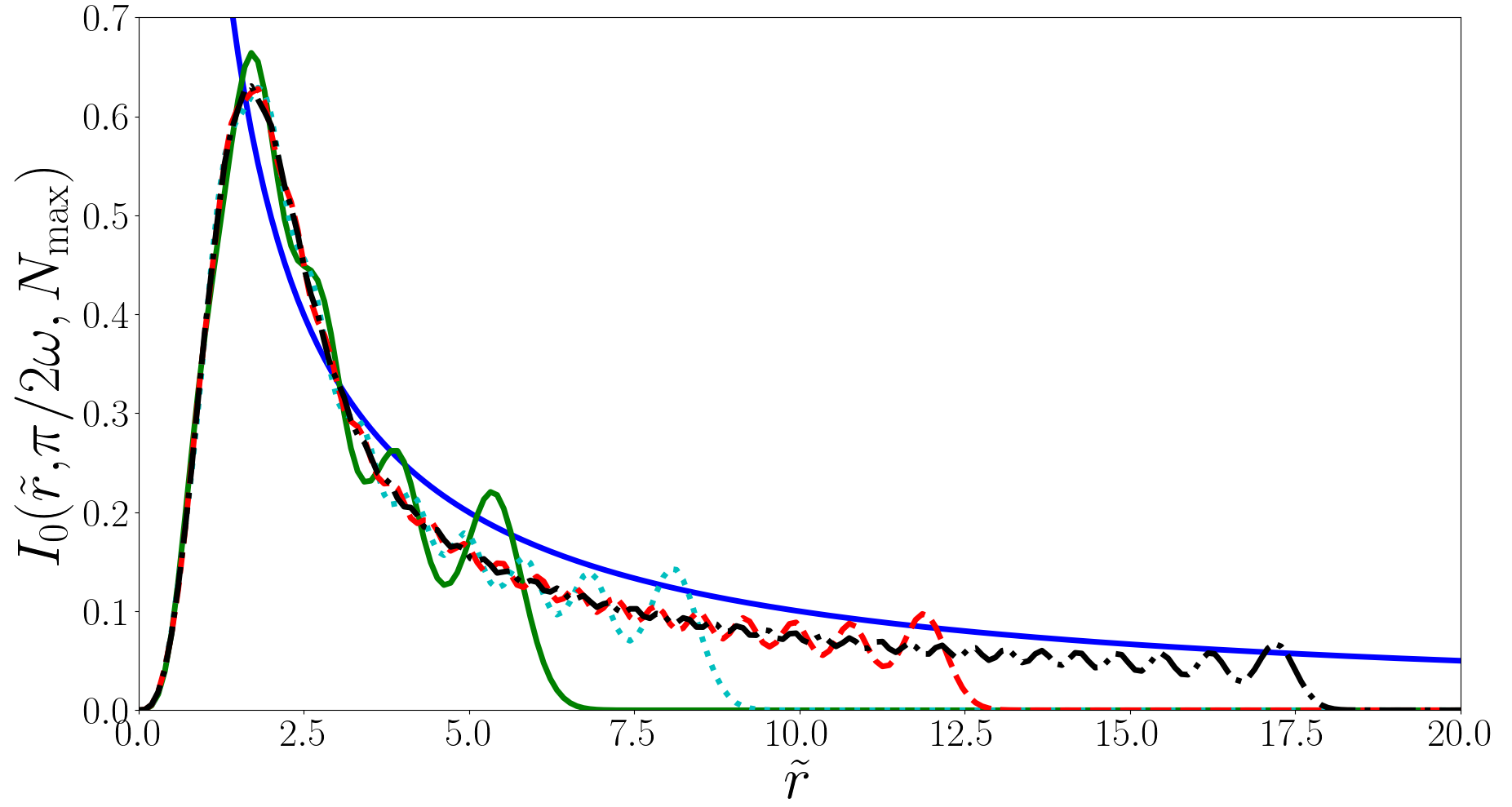}
\caption{$I_{0}({\tilde r},t=\pi/2\omega,N_{\rm max})$ as a function of ${\tilde r}=r/a_{\rm ho}$ for $N_{\rm max}=10$ (solid green curve),  $N_{\rm max}=20$ (dotted cyan curve), $N_{\rm max}=40$ (dashed red curve) and $N_{\rm max}=80$ (black dot-dashed curve). The thin solid blue curve is $1/\tilde{r}$.}
\label{fig:density}
\end{figure}

In Fig. \ref{fig:density} $I_{0}({\tilde r},t=\pi/2\omega,N_{\rm max})$ is plotted for increasing values of the cut-off ($N_{\rm max}$) in the sum over $m$ and $l$.  From this it is clear that as $N_{\rm max}$ increases there is $1/\tilde{r}$ tail in $I_{0}({\tilde r},t=\pi/2\omega,N)$. A similar analysis for $t=k\pi/\omega$, where $k$ is an integer, reveals that at these specific times $I_{0}({\tilde r},t=k \pi/\omega,N_{\rm max})$ does not exhibit a $1/\tilde{r}$ tail and hence the sum is convergent. However, away from $t=k\pi/\omega$ it is this $1/\tilde{r}$ dependence which leads to the divergence in $\langle \tilde{r}(t) \rangle$.

This divergence naturally leads to three questions: where does it come from, what does it mean in the context of experiment and why does this not happen in the reverse case? Answering the second question first: in the context of an experiment this result shows that if the system started in the ground state, with $a_{\rm s}/a_{\rm ho} \rightarrow \pm \infty$, and is quenched to the non-interacting regime an ensemble measurement of the separation between the two particles would exhibit a periodic structure but would also, away from $t=k \pi/\omega$, diverge. This is concerning on many levels and as such it is worth considering the first question which can be addressed by evaluating the wavefunction of the initial state (Eq.~(\ref{eq:psi1})), which in the limit  $\tilde{r}\rightarrow 0$ approaches $1/\tilde{r}$. A careful analysis of this reveals that completing the sum to $\infty$ in Eq.~(\ref{eq:psi1}) is crucial in describing the properties of the interacting wavefunction as $\tilde{r}\rightarrow 0$.  This means that after the quench has occurred (to the non-interacting regime) this cusp leads to a divergence in momenta which can only be transferred into high-energy non-interacting states. Since this work considers a zero-range interaction this means a divergence in momenta and hence a $1/\tilde{r}$ dependence in $I_0({\tilde r},t,N \rightarrow \infty)$. This then helps to understand what this means in the context of an actual experiment. In reality there is a cut-off, the interaction is not zero-range. However, this cut-off is short range and can be estimated to be of order  $<10^{-9}$m and hence one expects that the oscillations in $\langle \tilde{r}(t=\pi/2\omega) \rangle$ should have an amplitude of order $>7a_{\rm ho}$. This is still a very large oscillation, an order of magnitude larger than the case where the interactions are quenched from $a_{\rm s}/a_{\rm ho} \rightarrow 0$ to $a^{\prime}_{\rm s}/a_{\rm ho} \rightarrow \pm \infty$, see Fig.~\ref{fig:rt1}. For the third question; in the non-interacting to unitary case we believe there is no divergence because the wavefunctions being projected onto (the non-interacting states) do not have a cusp at $\tilde{r}=0$ and so there is no corresponding divergence in momenta.

\section{Conclusion}
In this work we have considered the dynamical properties of a harmonically trapped interacting two-body system, where the contact interactions are quenched between two values. Our work has focused on determining the Ramsey signal and the expectation value of the inter-particle separation particle. Although the methodology in determining the dynamical properties of these quantities is general in this work we have chosen to focus on two quenches: (i) from a non-interacting ($a_{\rm s}/a_{\rm ho} \rightarrow 0$) state to a unitary ($a^{\prime}_{\rm s}/a_{\rm ho} \rightarrow \pm \infty$) regime and (ii) from a unitary ($a_{\rm s}/a_{\rm ho} \rightarrow \pm \infty$) state to a non-interacting ($a^{\prime}_{\rm s}/a_{\rm ho} \rightarrow 0$) regime. In these scenarios results for the Ramsey signal are exact closed form expressions, and when the system is initially in the ground state the Ramsey signal is given by elementary functions.

For the Ramsey signal calculations we found periodic behaviour, as one would expect, however, it is notable that for the case where the quench is from the non-interacting state to the unitary regime the results on short-time scales match closely previous theoretical work in the limit of a single impurity in a uniform Fermi sea \cite{parish2016quantum}. This originates from the feature that the short term behaviour of these systems is governed by the two-body dynamics, as found in Refs. \cite{parish2016quantum,sykes2013quenching,corson2015bound} and is therefore independent of the trapping potential.

For the calculations of the expectation value of the separation between the two particles the results for a quenches between non-interacting and unitary regimes exhibited expected periodic behaviour with an amplitude of $\sim 0.3a_{\rm ho}$, for quenches from the non-interacting to the unitary regimes. For quenches from the unitary regime to the non-interacting regime we find a logarithmic divergence in the separation, which arises due to the contact interaction breaking down at small length scales. By imposing a cut-off based on the length scale of the Van der Waals force we expect that the size of these oscillations can be considerably larger $\sim 7a_{\rm ho}$. Additionally given that the $7a_{\rm ho}$ result is dependent on the interaction cut-off length scale this could open an avenue of investigation into the length scale of the inter-atomic interaction.

Finally we note that given experimental advances in the {\it building} of trapped few-atom systems \cite{serwane2011deterministic, murmann2015two, zurn2013pairing, zurn2012fermionization} it should possible to experimentally investigate the dynamics discussed in this work. Specifically   Ref. \cite{guan2019density} performed an experiment in which two trapped \textsuperscript{6}Li in distinct hyperfine states underwent a quench in trap geometry and the inter-atomic distant was measured. This is tantalizingly close to an experimental test of the results presented here, rather than a quench in trap geometry what is needed is a quench in interaction strength, which can and has been performed in other experiments \cite{cetina2016ultrafast}. Additionally,  we note that from the theoretical perspective this work can be extended to how the contact parameter \cite{tan2008energetics, tan2008generalized, tan2008large} which evolves as the two-body system is quenched and to the three-body problem \cite{fedorov2001regularization, PhysRevA.74.053604, levinsen2017universality, werner2006unitary, PhysRevA.76.033611, d2018efimov}.

\section{Acknowledgements}
 A.D.K. is supported by an Australian Government Research Training Program Scholarship and by the University of Melbourne.

\bibliographystyle{apsrev4-1}{}
\bibliography{Few-Body-Paper-submit}


\end{document}